\documentstyle[epsfig]{aipproc}

\newcommand{\beq}[1]{\begin{equation} \label{#1}}
\newcommand{\eeq}{\end{equation}}
\newcommand{\beqar}[1]{\begin{eqnarray} \label{#1}}
\newcommand{\eeqar}{\end{eqnarray}}                                            
\newcommand{\nid} {\noindent}
\newcommand{\bit}{\begin{itemize}}
\newcommand{\eit}{\end{itemize}}                    
\newcommand{\etc}{{\it et al }}
\newcommand{\htabt}{\hspace*{2mm}}
\newcommand{\bib}{\bibitem}                 
\newcommand{\npb}[3]{{\it Nucl. Phys.}~{\bf B#1,}~{#3}{(19#2)}} 
\newcommand{\prd}[3]{{\it Phys. Rev.}~{\bf D#1,}~{#3}{(19#2)}}
\newcommand{\prl}[3]{{\it Phys. Rev. Lett.}~{\bf #1,}~{#3}{(19#2)}}
\newcommand{\phlt}[3]{{\it Phys. Lett.}~{\bf B#1,}~{#3}{(19#2)}}

\begin{document}
\title{Deeply Virtual Compton Scattering\\ and Skewed Parton 
Distribution}

\author{Zhang Chen\thanks{ {\it email:} chenz@phys.mville.edu}\\[-5mm]}
\address{Department of Physics, Manhattanville College\\
Purchase, NY 10577\\[-5mm]}

\maketitle

\begin{abstract}
An overview of the current status of and possible future theoretical, 
phenomenological and experimental studies of DVCS and SPD's is presented. 
\end{abstract}

\section*{Introduction}

The study of the structure of the nucleon is one of the most important
frontiers in strong interaction physics. There are still
many unanswered questions largely due to the non-perturbative nature
of the bound state problem in QCD.  
Two traditional types of observables have been
studied extensively, both theoretically and experimentally, for the
last forty years: the parton (quark and gluon) distribution functions 
(PDF's) (via deeply inelastic scattering (DIS) or Drell-Yan processes),
and elastic form factors of the nucleon.
In the past few years, studies of a new type of nucleon observable,
the skewed parton distributions (SPD's), have 
flourished(eg,\cite{spd}). The SPD's generalize and interpolate between the ordinary
PDF's and elastic form factors and therefore contain rich structural
information. They can be measured in (exclusive) diffractive processes
in which the nucleon recoils elastically after receiving a non-zero
momentum transfer in the so-called deeply virtual 
limit\footnote{$Q^2 \!\rightarrow\!\infty$ while keeping Bjorken variable 
$x \!=\! x_B\! =\! \frac {Q^2} {2 p \cdot q}$ fixed, and 
$Q^2\!>\!>\!t\!=\! - (p'\!-\!p)^2$ 
where $p$ ($p'$) is the initial(recoil) nucleon momentum.}.

\section*{Theory: From DIS \& PDF's to DVCS \& SPD's}

Ordinary  PDF's in DIS are accessed through an optical theorem that relates the
imaginary part of a forward (Compton) scattering amplitude to the 
cross section. 
Through operator product expansion (OPE) one factorizes 
the hard scattering from the soft physics. 
The hard scattering 
can be calculated order by order in perturbation theory while the 
soft part is parametrized as PDF's. They are essentially
matrix elements of light-cone bilocal (quark and gluon) operators
between equal momentum (symmetric) states.
The factorization scale dependence of the matrix elements/PDF's is 
governed by a
renormalization group equation. The cross section is related to 
(usually a linear combination of) PDF's.  

\begin{figure}
\vspace*{-0.5cm}
\centerline{\epsfig{file=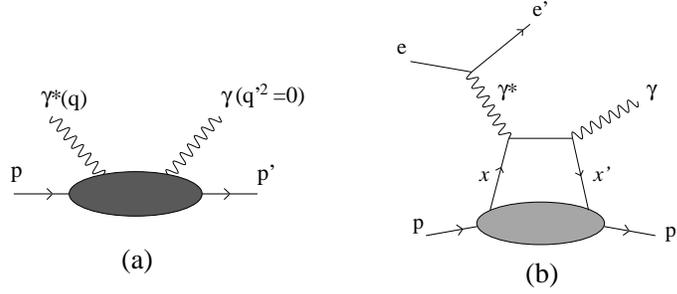,width=9cm}}
\vspace{4pt}
\caption{The DVCS process}
\label{fig:dvcs}
\end{figure}

The SPD's are, on the other hand, essentially matrix elements of the
same light-cone bilocal operators between {\it different} momentum
(asymmetric) states. They can be accessed thus via deeply virtual 
processes like diffractive vector meson production 
($ep \rightarrow 
ep' \,V\!M$)\cite{BrodskyVM,Radyphlt962,Collins&FS97,Frankf&Strikman98} 
and deeply virtual Compton scattering (DVCS)\cite{Dmuel,Ji97,Radyphlt961}. 
DVCS(See fig.\ref{fig:dvcs}. (b) is the lowest order
``handbag'' diagram in an e-p collision, where $x$ and $x'$
denote the longitudinal
momentum fractions of the interacting parton,) is a non-forward
process signified by non-zero $t$ with the longitudinal momentum
(fraction) transfer $x \!-\! x' \equiv \zeta \!=\! x_B$  and also 
in general a non-zero transverse momentum transfer. 

\begin{figure}[h]
\centerline{\epsfig{file=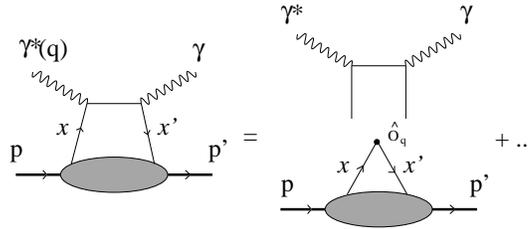,width=7cm}}
\vspace{4pt}
\caption{Non-forward OPE (only the lowest $q-q$ diagram is shown)}
\label{fig:ope}
\end{figure}

One still has valid factorization theorem\cite{Collins&FS97,Factorization} 
and OPE\cite{Dmuel,nfvcs} (fig.\ref{fig:ope}) in
the non-forward case. That is, the DVCS amplitude can be factorized
into a hard scattering part calculable perturbatively (the upper part
of the OPE diagram and also a crossed term) and a soft part (lower 
part of the diagram) parametrized as SPD's.  

SPD's, like usual PDF's, depend on factorization scale. Its QCD
evolution is governed again by a renormalization group equation
through a kernel that has been worked out to next-to-leading-order
(NLO)\cite{Belitsky}.\footnote{Also worked out is the NLO correction to
hard scattering in DVCS.} The QCD $Q^2$ evolution of the SPD's has been
studied extensively and has been shown to exhibit characteristics of 
both the 
DGLAP evolution of usual PDF's and the 
ERBL evolution of meson distribution amplitudes, depending on
different kinematic regions(eg, \cite{spd,nfvcs}). 
In particular, in the small $x_B$, ie, 
small $\xi \!=\! \frac {x_B} {2 (1 - x_B/2)}$ region
SPD's are closely connected with usual PDF's\cite{Collins&FS97}.
For $x\!>\!>\!\xi$, SPD's $\approx$ PDF's, while for $x \!\sim\! \xi$,  
in leading $\log {1 \over x} \sim \log {1 \over \xi}$, 
SPD's $\!\rightarrow\!$ forward PDF's\cite{BrodskyVM} and at large
factorization scale the $Q^2$ evolution tends to wash out effects of
the asymmetry $\xi$ (eg, \cite{Martin}). The region $x\!<\!\xi$ is 
least well-known. 

SPD's are indeed form
factors of the non-forward scattering amplitude
$\gamma^* p \rightarrow \gamma p'$, eg, take quarks (similar for
gluons)
\begin{eqnarray}
FT \,\, && \langle p',s'|\overline\psi(0)\gamma^\mu
			    \psi(z)|p,s \rangle
= H(x,\xi,t)\
	\overline u(p', s')\gamma^+ u(p,s) \nonumber \\
&&  \;\;\;\;\;\;\;\;\;\; + \; E(x,\xi,t)\
	\overline u(p', s'){i\sigma^{+\nu} (p'-p)_{\nu} \over 2M} u(p,s)
	+ \cdots  \ ,   \;\;  {\rm (Quark\; Spin\; Sum)} \nonumber    \\
%
FT \,\, &&  \langle p', s'|\overline\psi(0)\gamma^+\gamma_5
                            \psi(z)|p,s \rangle
= \widetilde{H}(x,\xi,t)\
        \overline u(p', s')\gamma^+ \gamma_5 u(p,s) \nonumber \\
&& \;\;\;\;\;\;\;\;\;\;\;\;  + \; \widetilde{E}(x,\xi,t)\
        \overline u(p',s') {(p'-p)^+ \gamma_5 \over 2M} u(p,s)
        + \cdots \ , \;\;\;\;\;  {\rm (Difference)} \nonumber
\end{eqnarray}
\nid where $H(x,\xi,t)$, $E(x,\xi,t)$,
$\widetilde{H}(x,\xi,t)$, $\widetilde{E}(x,\xi,t)$ are skewed quark
distributions with $ 2 \xi \!=\! \frac {x_B} {1 - x_B/2} $ 
\footnote{Notations in the literature differ. That of Ji's is 
used and mom. fractions
refer to ${1 \over 2}(p\!+\!p')$.}
and $FT$ denotes Fourier Transform.

While in general two SPDs correspond to one usual PDF, in the forward
limit of $p\!=\!p'$ ($\xi \!\rightarrow\! 0$ and $t \!\rightarrow\! 0$), 
SPD's are reduced to the normal distributions: 
$H(x,0,0) = q(x),~\widetilde{H}(x,0,0) = \Delta q(x),$ where $q(x)$ 
and $\Delta q(x)$  are the conventional forward quark and quark 
helicity distributions (Similar
equations hold for gluons). At the same time, the first moments of
these SPD's are related to nucleon form factors of corresponding EM or
Axial currents by the sum rules:
\begin{eqnarray}
\int^1_{-1} dx H(x,\xi,t) = F_1(t) \ , &\;\;&
\int^1_{-1} dx E(x,\xi,t) = F_2(t) \ ,     \nonumber     \\
\int^1_{-1} dx \widetilde{H}(x,\xi,t) = G_A(t)\ , &\;\;&
\int^1_{-1} dx \widetilde{E}(x,\xi,t) = G_P(t)\   \nonumber.
\end{eqnarray}
\nid At the same time, SPD's have many new features, in contrast to the
usual PDF's. 

\bit

\item{The non-forward amplitude to lowest $O(\alpha_s)$ is generally
related to a integration over the SPD's of type 
$Amp \!\sim\! \int \, dx \frac {1} {x - \zeta + i \epsilon}
f(x,\zeta, t)$ ($f$ denotes SPD's).}

\item{Cross section is obtained by squaring the 
amplitude. There is not an optical theorem, nor a simple
probability interpretation. SPD's can be viewed as overlap of 
wavefunctions between different parton numbers 
(Fock states)\cite{Fock}}

\item{SPD's are interference/correlation
functions of different wave functions/
probability amplitudes.
The extreme case with  $x \!>\! 0$  and  $x' \!<\! 0$ 
can be understood by re-interpret  $x'$ line as antiparton with 
momentum fraction $-x'$ resulting in ``extracting $q \overline{q}$-pair'' 
from the nucleon.}
\eit

SPD(via DVCS) is the only place so far probing the orbital angular 
momentum of partons\cite{Ji97}.
Moments of the SPD's provide information on quark and gluon
contributions to nucleon's spin because they are closely related to
the form factors of the (QCD) energy-momentum tensor. For example, one
can measure the skewed quark distribution in spin-averaged experiments
and extract form factors of the tensor. Extrapolating
to $t=0$ one can obtain the total (spin+orbital) quark contribution to
the nucleon spin $J_q(0)$ by taking $t=0$ in Ji's sum rule:
$$
{1 \over 2} \int \; dx \; x \; (H_q(x,\xi,t) + E_q(x,\xi,t)) = J_q(t).
$$
\nid There is a similar sum rule for gluon and one finds 
$J_q(0) + J_g(0) = {1 \over 2}$.

\section*{Phenomenology and Experiments}


Using known usual PDF's as input, models of SPD's have been studies by
making ansatz that fulfills the (some of which very non-trivial)
general properties of SPD's(eg, \cite{spd,models}). There are also
studies on contributions from meson exchange in the ERBL region 
($x\!>\!0$,$x'\!<\!0$, on SPD's as overlap of proton wavefunctions for
$x\!>\!\xi$\cite{Diehl} and on different dynamical models of SPD's,
including bag model, instanton vacuum and quark soliton model 
(eg, \cite{dmodels}).


Experiments usually access integrals of SPD's with parton momenta that 
are multi-variable and can not be directly obtained
from cross sections. Thus two main issues remain, one being experimental
difficulty, the other extracting SPD's from data.

The key background to DVCS is  
the QED Bethe-Heitler(BH) process(Compton scattering), which depend on
EM form factor, and its interference with DVCS. At higher $Q^2$ the
background is smaller, but so is the signal. This makes measuring DVCS
cross section very difficult experimentally. The first experimental
measurement come from HERA\cite{hera}, where the preliminary on 
96/97 data has shown evidence of the DVCS signal. However, the
interference also makes possible exploring DVCS at the amplitude level,
measuring its imaginary and real part independently.\footnote{A proposal for
JLAB Hall A\cite{jlab} proposed measuring cross section difference for
leptons of opposite helicities, which is proportional to the
interference of the imaginary part of the DVCS amplitude with a known
BH weight. This quantity turns out to be a {\sl linear} combination of
SPD's, thus if scaling is reached at the kinematic region with $6 Gev$
beam, it would be a measurement of SPDs' contribution. Also since the
higher twist effects is only down ${1 \over Q}$ and has a different
angular distribution (thus not masked by
leading-twist) it can give an estimation of these effects.}
There has been a lot of data on vector meson production
(eg, \cite{hermes,heravm}), but the extraction of SPD's is still a
pending task.\footnote{Another reason of the interest in exclusive diffractive
VM at small-x is that the cross section is predicted to be
proportional to the {\sl square} of gluon density\cite{BrodskyVM}.
This could be a direct measurement of the glue in the
proton, rather than getting it from evolution as in DIS.}
Table \ref{expov} is a very brief overview of the current experimental status.

\begin{table}[ht]
\caption{Overview of Status of SPD-related Exps, DVCS and
VM Production}
\label{expov}
\begin{tabular}{||c|c|c||}
\hline\hline
Process(es) & DATA & Proposed \\
\hline\hline
DVCS & HERA\cite{hera} ($e^+p \rightarrow e^+ \gamma p'$) 
& COMPASS, HERMES, JLAB\cite{jlab} \\
\hline
Meson Prod. & HERMES\cite{hermes}, HERA\cite{heravm}
& JLAB, COMPASS\\
\hline\hline
\end{tabular}
\vspace*{-0.4cm}
\end{table}



Independent of the actual form of SPD's, factorization predicts 
that in the scaling limit of $Q^2 \!\rightarrow\! \infty; x_B, t$
fixed, DVCS $\!\sim\! {1 \over Q^0}$ and VM production 
$\!\sim\! {1 \over Q}$. There are also helicity selection rules 
stating that for DVCS only 
$\gamma^* (T) \!\rightarrow\! \gamma (T) \!\sim\! {1 \over Q^0}$ 
and $\gamma^* (L) \!\rightarrow\! \gamma (T)$  
is power suppressed while for VM only 
$\gamma^* (L) \!\rightarrow\! VM (L) \!\sim\! {1 \over Q}$ and all else are
power suppressed. This is due to helicity conservation and introduces
helicity parton distributions (eg, helicity flip gluon distribution
that has no counterpart in DIS)\cite{helipdf}, while being
higher-twist gives the suppression. Therefore by looking at the
angular distributions one have a window on higher-twist effects
\cite{htwist} since it is not masked by leading twist because of 
different helicity.

Other SPD-related experimental processes been 
proposed include diffractive dijets (photo-production) and crossed DVCS 
(eg, $\gamma^* \gamma \!\rightarrow\!\pi^+\pi_-$)\cite{other}, etc. 

\section*{Acknowledgements}

I thank Xiangdong Ji and Macus Diehl for helpful discussions. This
research is partly supported by a grant from Manhattanville College.

\end{document}